# Graphical Abstract

**BalCon – resource balancing algorithm for VM consolidation**

Andrei Gudkov, Pavel Popov, Stepan Romanov

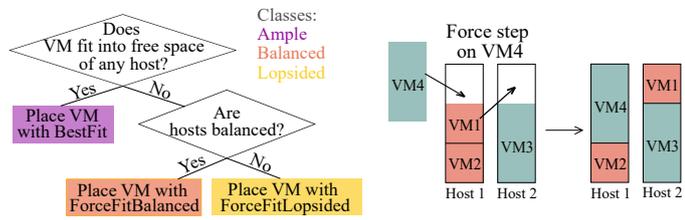

# Highlights

## BalCon – resource balancing algorithm for VM consolidation

Andrei Gudkov , Pavel Popov , Stepan Romanov

- We introduce the BalCon algorithm for solving the migration-aware consolidation problem. Force Fit is a key part of BalCon that allows for solving instances with Force Steps.
- Huawei and synthetic datasets are used to validate BalCon.
- Integer linear programming flavor flow model is proposed to produce optimal solutions and lower bounds.
- A Sercon heuristic is modified for comparison with BalCon.
- The balance factor is proposed as a measure of free space distribution in a datacenter.

# BalCon – resource balancing algorithm for VM consolidation

Andrei Gudkov[a], Pavel Popov[a], Stepan Romanov[a]

[a]*Huawei Technologies Company Ltd, Russian Research Institute, Moscow, 121099, Russian Federation*

## Abstract

Cloud providers handle substantial number of requests to create and delete virtual machines (VMs) on a daily basis, where the unknown sequence of requests eventually leads to resource fragmentation. To mitigate this issue, periodic consolidation of VMs into fewer number of physical hosts is an important cost-saving procedure, closely related to the vector bin-packing problem. In this paper, we propose the BalCon algorithm for consolidation that steadily reduces the number of active hosts and keeps migration costs low. BalCon classifies the cluster's state and selects one of three heuristics to balance resources for superior consolidation. To evaluate BalCon's performance with respect to optimality, we introduce integer programming models. BalCon finds 99.7% of the optimal solutions for over 750 problem instances. This outstanding result was achieved due to the Force Step of our algorithm, which is the key improvement detail for common heuristics. We compare BalCon with a modified Sercon heuristic using Huawei and synthetic datasets with two resources for allocation.

*Keywords:* Consolidation, Bin packing, High-multiplicity, Cloud computing, Resource scheduling, Heuristic algorithms

## 1. Introduction

Virtualization technology allows the sharing of a physical machine (host) between multiple isolated virtual machines (VMs) [1, 2]. Resources for virtualization include the central processing unit (CPU), random access memory (RAM), network, drives, *etc.* Several users can run VMs with different operating system (OS) on the same host simultaneously. The resource sharing prospect of virtualization opens up variety of opportunities for business.

Cloud providers operate datacenters with thousands of hosts and exploit virtualization to lease VMs to clients [3, 4]. Resources combinations for the VMs are usually predefined and known as flavors. For instance, two dimensional flavors (CPU, RAM) can be (2 cores, 4 GiB), (8 cores, 24 GiB), (64 cores, 560 GiB), *etc*. Clients choose a flavor and an OS image for the VM and remotely use the VM on a "pay-as-you-go" basis. The guarantees from providers to clients regarding VMs availability and reliability are specified in a service level agreement (SLA) [5, 6], where violation leads to the providers being penalised. Therefore, during datacenters operation the providers strive to minimize SLA violations and operational costs.

Hosts in an idle state consume about 70% of maximal power consumption[7]. To minimize operational costs, providers seek to allocate VMs into the fewest number of hosts and power off the unused ones. The scope for optimization arises from the uncertainty of requests during cloud operation, which leads to fragmentation of resources.

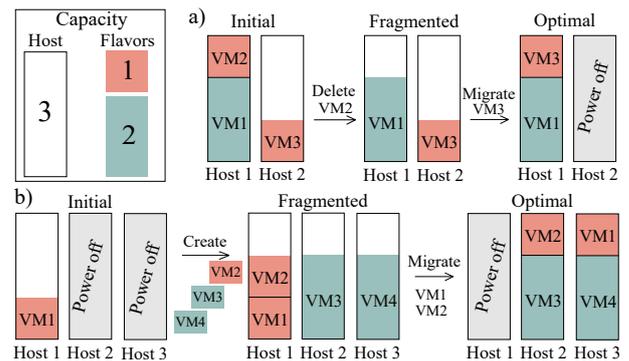

Figure 1: An example of resource fragmentation caused by delete and create operations followed by optimisation via the migration operation. Panel a) demonstrates how a delete operation of VM2 led to resource fragmentation, which was optimized by migration of VM3. Panel b) shows how a sequence of create operation for the trace of VMs (VM2, VM3, VM4) by the First Fit heuristic caused resource fragmentation, which was optimized by the migration of VM1 and VM2.

The VMs in datacenters are managed through three essential operations: 1) create VM on host, 2) delete VM from host, and 3) migrate VM from one host to another. The periodic deletion of VMs (see Figure 1a) along with unpredictable requests to create VMs (see Figure 1b) leads to resource fragmentation on the hosts and non-optimal VM placement.

Live migration is an important technique for the optimization of datacenters, as it allows running VMs to be migrated between hosts without interruption or client notice [8]. However, frequent live migration is undesirable due to the increased load on the management side of the cloud, where the load depends on various factors, such as the state of the VMs (processes, memory, network, *etc.*). Therefore, it is preferable to migrate less active VMs with less RAM.

*1.1. Consolidation*

Consolidation of virtual machines [9] is a procedure that searches for a better assignment of VMs to hosts. The improved assignment benefits one or more of the following metrics: power consumption [10, 7], number of active hosts [11, 12], migration cost [13, 11, 14, 15], resource fragmentation [12, 16], number of SLA violations [11], *etc*. Typically, minimization of the number of active hosts is a primary goal of consolidation, however taking migration cost into account is also crucial. Indeed, large number of migrations introduce undesirable loads on the management side of a datacenter. To avoid overloads, a balance between migration cost and the number of active hosts must be preserved.

Consolidation relies on two major approaches: static and dynamic [9]. Static consolidation reserves resources for a VM based on its maximal consumption, which is specified in the flavor. Dynamic consolidation assigns VMs to hosts using either their current or predicted, by statistical model, resource requirements of VMs. These requirements are usually lower than the flavor size. In other words, static consolidation guarantees the availability of the resources on the host for the clients, whereas dynamic consolidation provides denser packing at the cost of possible SLA violations – if demand of VMs on a host rises to a higher level.

Furthermore, algorithms for consolidation are divided into centralized [9, 10] and decentralized [17, 18, 19] depending on the cloud architecture. In centralized algorithms, any host possesses full information about the entire cluster. In decentralized algorithms, all hosts pursue the same goal, however each host has information only about its neighborhood. Consolidation for centralized architectures works faster and provides better results, whereas decentralized architectures are free from a single point of failure and have better scalability.

*1.2. d-dimensional vector bin packing*

Consolidation of VMs is closely related to the well-known *d*-dimensional vector bin packing (d-VBP) problem. In the d-VBP problem, items are represented as d-dimensional vectors which are packed into the fewest number of d-dimensional bins such that: 1) all items are placed 2) the sum of items in each bin is less than or equal to the bin capacity in all dimensions. The d-VBP problem is known to be an NP-hard problem which is exponentially hard to solve [20]. However, if $d = 1$, there are two types of restrictions of the d-VBP problem that are often applicable to consolidation and lead to optimal solutions in polynomial time: 1) divisible items – in a list of sorted items each item is a divisor of next items – can be optimally packed with the First Fit Decreasing heuristic [21], 2) high-multiplicity VBP problem – items form a few groups and the items in each group are identical – can be solved in polynomial time [22]. To the best of our knowledge, the d-VBP problem for $d \neq 1$ is still exponentially complex even with the above restrictions.

The hardness of the d-VBP problem requires the use of heuristics to obtain a solution in reasonable time [23, 24, 25]. Unfortunately, the d-VBP misses objective of minimizing of migration cost, which restricts its application to migration-aware consolidation [14]. Application of d-VBP heuristics to the consolidation problem is equivalent to removing all the VMs from the hosts and placing them again. Such actions require unacceptably large amounts of memory for migration.

*1.3. Methods for solving the consolidation problem and proposed approach*

To solve the multidimensional consolidation problems researchers mainly utilize three methods: 1) integer linear programming (ILP), 2) metaheuristics, and 3) heuristics. ILP formulations obtain the optimal solution only for the small problem instances because of the exponential time complexity of solvers [26, 27, 28]. Metaheuristics work faster than ILP solvers and provide approximate solutions for larger instances. Such metaheuristics algorithms include: genetic algorithms[29, 30], ant colony optimization [16, 31], particle swarm optimization [32], artificial bee colony [33], *etc.* Typically, the performance of metaheuristics is worse than that of the heuristics in terms of the solvable instance size, parameters dependence, and time complexity[34].

Heuristics for the consolidation problem practically used due to time efficiency, small number of parameters, and implementation simplicity. The majority of migration-aware heuristics rely on this common approach: sort hosts, sort VMs, and try to place largest VM into the most loaded host [14, 11, 35, 12]. If the VM does not fit into a host because of a lack of free space, then the heuristic skips the host and try the next one. We call such skipping heuristics that exploit only available free space – "Sercon-like". The heuristics differ in the criteria for sorting of d-dimensional objects, additional classifications, and algorithm parameters. *Murtazaev* and *Oh* performed sorting using "surrogate weight" and introduced parameters for the maximum number of migrations and minimum migration efficiency [14]. *Ferreto et al.* used lexicographic order for sorting and held steady VMs[11]. *Rao* and *Thilagam* proposed hosts classification into receivers and donors based on the theoretical minimum of required hosts and introduced a defragmentation procedure to reduce resource fragmentation after consolidation [12].

The main disadvantage of Sercon-like heuristics becomes evident in consolidation problems with 2 resources. The heuristics are poorly adapted to solving instances with imbalanced hosts lacking free space for one of the resources (see Figure 2a, RAM limits Host 2, whereas CPU limits Host 3). Sercon-like heuristics are unable to free any of the hosts from the assignment demonstrated in Figure 2a. However, migrations of yellow and green VMs between Host 2 and Host 3 increase free space and release Host 1 (see Figure 2b).

In this paper, we propose the BalCon algorithm for solving 2D centralized consolidation problems with a controlled trade-off between the number of released hosts and the amount of migrated memory. The key component of BalCon is the Force Step, which places a VM into the host with a lack of free space using induced migrations. For instance, Figure 2c illustrates two Force Steps that are required to consolidate the initial assignment (Figure 2a) into the optimal assignment (Figure 2b). To efficiently place 2D VMs, we introduce the balance factor to classify the cluster state and choose one of three heuristic for VM placement. To evaluate BalCon we use two datasets: real

VM assignments from Huawei Cloud datacenters and synthetically generated hard instances for the purpose of stress testing. Also we introduce ILP models and a modified the Sercon heuristic for comparison. The results indicate the superior performance of the BalCon towards solving consolidation problems under resource imbalance. The code with the implementation of BalCon and the synthetic datasets are presented in https://github.com/andreigudkov/BalCon.

This paper is organized as follows. In Section 2, we formulate the cloud consolidation problem. In Section 3, we introduce integer programming models that were used to obtain optimal solutions and lower bounds for performance evaluation. In Section 4, we provide a description of the BalCon algorithm, including the basic ideas (as shown in the diagram in Section 4.1) and the formal pseudo code (as presented in Section 4.3). In Section 5, we describe the datasets and compare BalCon with Sercon and optimal solutions. In Section 6, we conclude the paper and propose possible directions for future work.

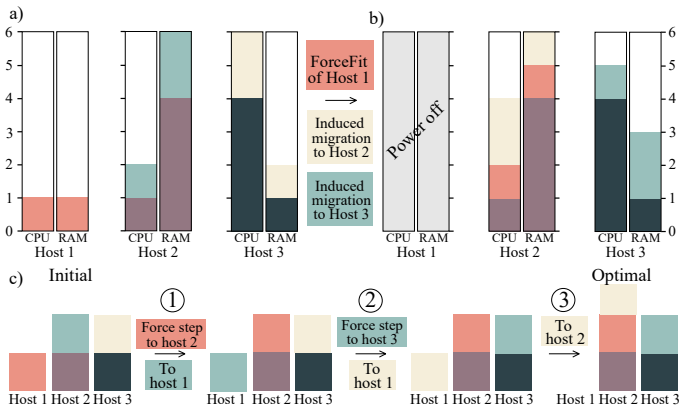

Figure 2: A consolidation problem instance with two resources (CPU, RAM) which is solvable by BalCon and unsolvable by Sercon-like heuristics. The problem involves four color-coded VMs that need to be assigned to three hosts, each with a capacity of (6, 6) *a.u.* Panel a) shows the initial assignment of VMs to hosts. Panel b) shows the optimal assignment of VMs to hosts. Panel c) shows the sequence of actions taken by the BalCon algorithm to optimize the initial assignment panel a) to the final assignment panel b).

## 2. Problem statement

In our scenario, we operate a cluster with sets of physical hosts $h \in H$, VMs $v \in V$, and flavors $f \in F$. Every host $h$ provides a pair of integer resources for scheduling: CPU (cores) and RAM (tebibytes). We denote the capacity of host $h$ as $(h.cpu, h.mem)$, and the capacity of other resources is denoted in the same way. The demand for resources of a VM $v$ is defined in a flavor $f$, that is, $(v.cpu, v.mem) = (f.cpu, f.mem)$. Each host $h$ reserves resources for the allocated VM $v$.

**Definition 1.** A mapping $\mu$ is the assignment of VMs to hosts, where some VMs can be unassigned. The mapping $\mu$ is feasible if:

1. every VM is assigned to exactly one host.

2. capacity of all hosts is satisfied in both (CPU and RAM) dimensions:
$\forall h \in H, \sum_{v \in h} v.cpu \leq h.cpu \wedge \sum_{v \in h} v.mem \leq h.mem$

Our goal is to consolidate the VMs of the cluster. Given an initial feasible mapping $\mu_0$ of the VMs to the hosts, we try to compute a different feasible mapping $\mu_b$ that minimizes the number of active hosts and the amount of migrated memory.

**Definition 2.** For a given mapping $\mu$, host $h$ is called *active* if it contains at least one VM $v$. $\exists v : \mu(v) = h$

**Definition 3.** For a given mapping $\mu$, a VM $v$ is called *migrated* if it changes host: $\mu(v) \neq \mu_0(v)$

To control our preference between the number of active hosts $A(\mu_b)$ and the amount of migrated RAM $M(\mu_b, \mu_0)$, we minimize the following linear objective function:

$$Obj(\mu_b, \mu_0) = w_a \cdot A(\mu_b) + w_m \cdot M(\mu_b, \mu_0) \quad (1)$$

where $w_a, w_m$ are non-negative weights. It is important to note that the resulting mapping of the minimization problem depends on the ratio of the weights instead of their absolute values. To investigate the algorithms performance under different objective functions, we introduce the maximal memory for migration per host

$$MPH = \frac{w_a}{w_m} \text{ TiB}. \quad (2)$$

The parameter is used for evaluation (as described in Section 5.2). Additionally, in our algorithm, if the migration cost to release any host exceeds $MPH$, then releasing the host decreases the objective function. In the case of the classical VBP problem, $MPH = \infty$ TiB.

## 3. Integer programming solutions

### 3.1. Allocation model

Bartók and Mann [26] presented a straightforward Integer Linear Programming (ILP) formulation of the similar consolidation problem. However, our problem statement differs in the calculation of migration cost. Their formulation counts the number of migrated VMs, whereas we compute the total amount of migrated memory. The updated ILP allocation model is expressed in our terms as follows.

The binary variables are:

$\text{Alloc}_{v,h} = 1$ if $v$ is allocated on $h$ and 0 otherwise
$\text{Active}_h = 1$ if $h$ is active and 0 otherwise
$\text{Migr}_v = 1$ if $v$ is migrated and 0 otherwise

With these variables the objective is to

$$\text{minimize } w_a \sum_{h \in H} \text{Active}_h + w_m \sum_{v \in V} v.mem \cdot \text{Migr}_v. \quad (3)$$

In other words, our objective is to minimize the number of active hosts and the amount of migrated memory, subject to the following constraints:

$$\sum_{h \in H} \mathsf{Alloc}_{v,h} = 1 \qquad \forall v \in V \qquad (4)$$

$$\sum_{v \in V} v.cpu \cdot \mathsf{Alloc}_{v,h} \leq h.cpu \qquad \forall h \in H \qquad (5)$$

$$\sum_{v \in V} v.mem \cdot \mathsf{Alloc}_{v,h} \leq h.mem \qquad \forall h \in H \qquad (6)$$

$$\mathsf{Alloc}_{v,h} \leq \mathsf{Active}_h \qquad \forall (v,h) \in V \times H \qquad (7)$$

$$\mathsf{Migr}_v = 1 - \mathsf{Alloc}_{v,\mu_0(v)} \qquad \forall v \in V \qquad (8)$$

Constraint 4 requires that every VM is scheduled on exactly one host (Definition 1.1). Resource Constraints 5 and 6 ensure that the CPU and RAM capacities of the hosts are satisfied (Definition 1.2). Constraint 7 guarantees that hosts with VMs are active (Definition 2). According to Constraint 8, a VM is migrated if it changes its host (Definition 3).

*3.2. Flavor flow model*

We introduce the flavor flow model that takes advantage of the fewer number of flavors than VMs. The allocation model uses $|H|$ variables for each VM, however many of the VMs are of the same flavor and have equal resource demands ($f.cpu$, $f.mem$). Permutation of these similar VMs leads to the same values of objective function and an undesirable symmetry. Instead of defining the exact location for every VM, we keep track on the flow of the VMs of the same flavor between hosts. Such a flavor flow model breaks the symmetry of the allocation model because all VMs of the same flavor are equal.

Formal definition of the flavor flow model requires defining the following variables:

$\mathsf{In}_{f,h} \in \mathbb{Z}$ is the number of VMs of flavor $f$ migrated into host $h$

$\mathsf{Out}_{f,h} \in \mathbb{Z}$ is the number of VMs of flavor $f$ migrated out from host $h$

$\mathsf{Active}_h = 1$ if $h$ is active and 0 otherwise

With these variables the objective is to

$$\text{minimize } w_a \sum_{h \in H} \mathsf{Active}_h + w_m \sum_{f \in F, h \in H} f.mem \cdot \mathsf{Out}_{f,h}. \qquad (9)$$

Let $n_{f,h}$ be the number of VMs of flavor $f$ initially located on host $h$, according to the mapping $\mu_0$. Then the constraints of the flavor flow model are:

$$\sum_{h \in H} \mathsf{Out}_{f,h} = \sum_{h \in H} \mathsf{In}_{f,h} \qquad \forall f \in F \qquad (10)$$

$$\mathsf{Out}_{f,h} \leq n_{f,h} \qquad \forall (f,h) \in F \times H \qquad (11)$$

$$n_{f,h} \cdot (1 - \mathsf{Active}_h) \leq \mathsf{Out}_{f,h} \qquad \forall (f,h) \in F \times H \qquad (12)$$

$$\sum_{f \in F} f.cpu \cdot (n_{f,h} + \mathsf{In}_{f,h} - \mathsf{Out}_{f,h}) \leq h.cpu \cdot \mathsf{Active}_h \quad \forall h \in H \quad (13)$$

$$\sum_{f \in F} f.mem \cdot (n_{f,h} + \mathsf{In}_{f,h} - \mathsf{Out}_{f,h}) \leq h.mem \cdot \mathsf{Active}_h \quad \forall h \in H \quad (14)$$

Table 1: Comparison of the models by domain of variables and number of constraints.

| Model | Domain of variables | Number of constraints |
|---|---|---|
| Allocation | $\mathbb{B}^{|V||H|+|V|+|H|}$ | $|V||H| + 2|H| + 2|V|$ |
| Flavor flow | $\mathbb{B}^{|H|} \times \mathbb{Z}^{2|F||H|}$ | $2|F||H| + 2|H| + |F|$ |
| Relaxed Flavor flow (Lower Bound) | $\mathbb{B}^{|H|} \times \mathbb{R}^{2|F||H|}$ | $2|F||H| + 2|H| + |F|$ |

Constraint 10 ensures that the net flow of VMs is zero for each flavor $f$. In other words, the number of VMs migrated out from all hosts must be equal to the number of VMs migrated onto all hosts for each flavor $f$. Constraint 11 forbids moving out of a host more VMs of a flavor $f$ than were initially present on the host. For inactive hosts, all VMs must be moved out according to Constraint 12. Resource Constraints 13 and 14 ensure that the CPU and RAM capacities of each host are satisfied.

The flavor flow model has asymptotically $|V|/|F|$ fewer variables than the allocation model, which allows for solving much larger problem instances when $|F| \ll |V|$ – a typical real cloud situation.

*3.3. Lower bounds*

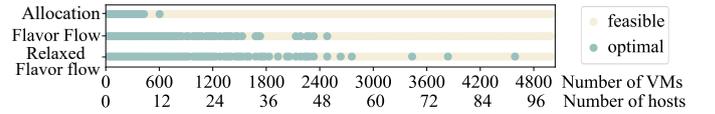

Figure 3: Comparison of the ILP models on progressively larger instance sizes. The problem instances of different sizes were solved by the CBC solver for 10 minutes each. The mean number of flavors per problem instance was 27.

We detected that the running time of ILP solvers is sensitive to the value of *MPH*. When *MPH* is close to 0 or infinity, solvers manage to find the optimal solutions $\mu_{opt}$ in a reasonable time. Unfortunately, for intermediate values of *MPH*, solvers are unable to provide the optimal solutions even in a few days. Nevertheless, relaxation of the flavor flow model allows one to obtain a lower bound (LB) $\mu_{LB}$ on the optimal solution for the intermediate *MPH* values. To perform the relaxation, we kept the $|H|$ variables $\mathsf{Active}_h$ as binary and relaxed the domain of the $2|F||H|$ variables $\mathsf{In}_{f,h}$ and $\mathsf{Out}_{f,h}$ to $\mathbb{R}$. This relaxation significantly simplifies the flow model for the solver and sets the lower bound on the optimal solution

$$Obj(\mu_{LB}, \mu_0) \leq Obj(\mu_{opt}, \mu_0). \qquad (15)$$

The comparison of the allocation, flavor flow, and relaxed flavor flow models in terms of variable domain and number of constraints is presented in Table 1. To compare the performance of the models, we searched for the optimal solution of problem instances with different sizes using the CBC solver [36] for 10 minutes (see Figure 3). As expected, the flavor flow models provided the optimal solutions for much larger instances than the allocation model.

## 4. Algorithms

### 4.1. Basic description of BalCon

The BalCon algorithm was created to cope with imbalanced situations in a datacenter. The need for such an algorithms arises from the poor performance of Sercon-like heuristics in terms of host balancing. For example, Sercon-like heuristics are unable to consolidate the imbalanced mapping from Figure 2a because of insufficient amount of free space in the host for one of the resources (RAM limits Host 2, whereas CPU limits Host 3). However, the consolidation is possible by shuffling the yellow and green VMs between Host 2 and Host 3 (see Figure 2b). The shuffling in the BalCon is implemented with Force Steps, which free space in a host by ejecting VMs from the host into a temporary buffer (stash $S$). The VMs from the stash are later moved into other hosts, which leads to induced migrations. To perform the consolidation in the example, BalCon uses two Force Steps. The sequence of actions for the consolidation is presented in Figure 2c, where Host 1 illustrates the role of the stash. In the first step, to place the red VM to Host 2 we eject the green VM from Host 2. In the second step, to place the green VM to Host 3 we eject the yellow VM from Host 3. Finally, we place the yellow VM into the free space of Host 2. As a result, we released Host 1 and induced two migrations: 1) the yellow VM was migrated from Host 3 to Host 2 and 2) the green VM was migrated from Host 2 to Host 3.

In general, BalCon is a greedy heuristic that attempts to release one active host in each step. The basic ideas of the algorithm are presented in Figure 4a, whereas a formal definition will be given in Section 4.3. In each iteration, BalCon chooses the smallest host according to migration cost and places a VM from the host into the stash $S$.

**Definition 4.** The migration cost of VM $v$ is $v.mem$, and the migration cost of host $h$ is $\sum_{w \in h} w.mem$, where $w$ are unmigrated VMs: $\mu_0(w) = h$.

**Definition 5.** The stash $S$ is a temporary buffer for VMs $v$ with a resource vector $s = (\sum_{v \in S} v.cpu, \sum_{v \in S} v.mem)$

For the success of a host release, $S$ must be emptied. The BalCon takes the largest VM from the stash and uses the Best Fit heuristic until it encounters a problem with free space on all the hosts. The largest VM $v$ is defined using the following formula:

$$|v| = \frac{v.cpu}{\sum_{w \in V} w.cpu} + \frac{v.mem}{\sum_{w \in V} w.mem}. \quad (16)$$

The free space of a host $h$ is defined by the mean of its load and allocated VMs $v \in h$

$$\text{load}(h) = \left( \sum_{v \in h} v.cpu, \sum_{v \in h} v.mem \right) \quad (17)$$

$$\text{free}(h) = (h.cpu - \text{load}(h).cpu, \ h.mem - \text{load}(h).mem) \quad (18)$$

In the case of a lack of free space on the hosts, BalCon classifies the situation using the balance factor (defined formally in Section 4.2) and chooses one of two heuristics with Force Steps: *ForceFitBalanced* or *ForceFitLopsided*. These heuristics pick a destination host $h$ and free space for a given VM $v$ by ejecting VMs from $h$ into the stash. This procedure repeats and VMs from the stash are assigned to new or their original hosts. The emptying of the stash indicates the generation of a new feasible mapping. In the new mapping some of the ejected VMs could migrate between active hosts.

**Definition 6.** A migration of a VM is called *induced* if it takes place between active hosts.

Using induced migrations BalCon solves the problem of deficiency of free space in a datacenter. Induced migrations allow one to shuffle VMs and release the host, at the cost of an increase in the amount of migrated memory.

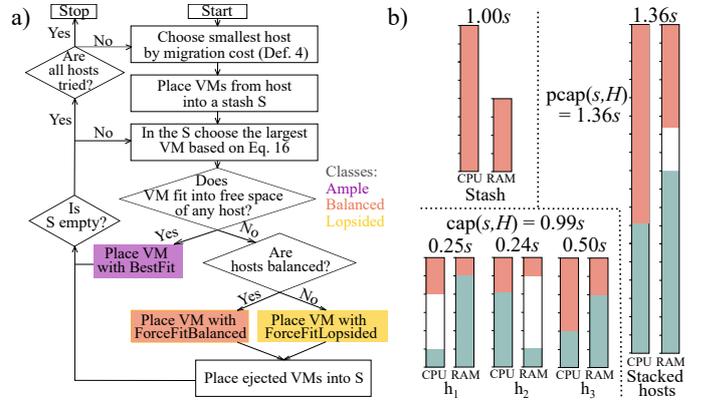

Figure 4: a) A diagram illustrates the workflow of the BalCon algorithm b) An illustration of the calculation of cap (Eq. 20) and pcap (Eq. 21) for the balance factor. The stash resource vector $s$ (red color) represents a unit vector to measure the amount of the free space. The green color corresponds to the occupied space in the hosts by VMs.

### 4.2. Balance factor

We introduce the balance factor as a measure of the distribution of free space in a datacenter. The factor allows us to estimate the potential for VM shuffling operation to increase VM placements. In each step, BalCon attempts to fit the stash $S$ into the hosts. Therefore, the distribution of free space between the hosts is improved if more stashes can be accommodated by the hosts. The balance factor is defined as the ratio of the current amount of stashes in the free space of the hosts $\text{cap}(s, H)$ to the potential amount of stashes in the combined free space of all hosts $\text{pcap}(s, H)$

$$BF(s, H) = \frac{\text{cap}(s, H)}{\text{pcap}(s, H)}, \quad (19)$$

where

$$\text{cap}(s, H) = \sum_{h \in H} \min \left( \frac{\text{free}(h).cpu}{s.cpu}, \frac{\text{free}(h).mem}{s.mem} \right) \quad (20)$$

$$\text{pcap}(s, H) = \min \left( \sum_{h \in H} \frac{\text{free}(h).cpu}{s.cpu}, \sum_{h \in H} \frac{\text{free}(h).mem}{s.mem} \right). \quad (21)$$

The domain of $BF(s, H)$ is $[0, 1]$, where the value of 1 corresponds to a purely balanced situation without potential for improvement, whereas a value of 0 corresponds to a purely lopsided situation that can be improved. In the balanced case most hosts reach their limit by the same resource, unlike in the lopsided case where some hosts are limited by *cpu* and some by *mem*.

Note that both $\mathsf{pcap}(s, H)$ and $\mathsf{cap}(s, H)$ provide relaxed values that ignore the actual sizes of separate VMs in the stash. Thus, $\mathsf{pcap}(s, H)$ and $\mathsf{cap}(s, H)$ set upper bounds on the amount of stashes in any mapping and current mapping respectively. For instance, if $\mathsf{cap}(s, H) < 1$ the allocation of all VMs from the stash is impossible without induced migrations. Similarly, if $\mathsf{pcap}(s, H) < 1$ the stash is implacable into the hosts by any mapping. To distinguish between Balanced and Lopsided situations, we introduce the parameter $\alpha$. If $BF(s, H) < \alpha$ or $\mathsf{cap}(s, H) < 1$ the situation is Lopsided and Balanced otherwise. In practice we use $\alpha = 0.95$.

An example of capacities calculation is presented in Figure 4b. The stash with a resource capacity of $s = (8.0, 4.0)$ is used as a unit vector of free space. Additionally, there are three partially filled hosts with equal capacities of $(6.0, 6.0)$ and free spaces $\mathsf{free}(h_1) = (5.0, 1.0)$, $\mathsf{free}(h_2) = (1.9, 5.0)$, and $\mathsf{free}(h_3) = (4.0, 2.0)$. In total, the hosts can allocate at most

$$\begin{aligned}\mathsf{cap}(s, H) &= \min\left(\frac{5.0}{8.0}, \frac{1.0}{4.0}\right) + \min\left(\frac{1.9}{8.0}, \frac{5.0}{4.0}\right) + \\ &+ \min\left(\frac{4.0}{8.0}, \frac{2.0}{4.0}\right) = 0.25 + 0.24 + 0.50 = 0.99\end{aligned} \quad (22)$$

of the stashes without induced migrations. Potential capacity is computed by combining the three hosts together that results in

$$\begin{aligned}\mathsf{pcap}(s, H) &= \min\left(\frac{5.0 + 1.9 + 40}{8.0}, \frac{1.0 + 5.0 + 2.0}{4.0}\right) = \\ &= 1.36\end{aligned} \quad (23)$$

of the stashes sizes. Finally, we compute the balance factor $BF(s, H) = \mathsf{cap}(H)/\mathsf{pcap}(H) = 0.73$. The fact that the BF is lower than $\alpha = 0.95$ indicates a resource imbalance and potential for improvement by induced migrations.

*4.3. The BalCon algorithm*

The core structure of our algorithm is presented in the Listing 1. There are three global parameters in the algorithm:

- $\alpha$ is used as a threshold to classify imbalanced $(BF(s, H) < \alpha)$ and balanced $(BF(s, H) \geq \alpha)$ situations using Eq. 19.

- $b$ is the maximal number of Force Steps to try during host release.

- $\gamma$ is the maximal number of tries of the same destination host in a row.

**Listing 1:** High-level BalCon algorithm structure

| | |
|---|---|
| **Input** | : $H$ is the list of hosts |
| | $V$ is the list of VMs |
| | $\mu_0$ is the initial feasible mapping |
| **Global Parameters:** | $\alpha = 0.95$ |
| | $b = 4000$ |
| | $\gamma = 3$ |
| **Output** | : At the end, $\mu_b$ is a feasible |
| | mapping of VMs to hosts |

1 **Procedure** `BalCon`($H$, $V$, $\mu_0$)
2     $\mu_b :=$ **copy** $\mu_0$
3     $H :=$ **sort** $H$ by migration cost
4     **for** $h$ in $H$ **do**
5        $\mu_{tmp} :=$ **copy** $\mu_b$
6        $S :=$ **get** VMs from $h$ according to $\mu_{tmp}$
7        $\mu_{tmp} :=$ **Unassign** all VMs in $S$ from $\mu_{tmp}$
8        $A :=$ **get** active hosts from $\mu_{tmp}$
9        $\mu_{tmp} :=$ `ForceFit`($S$, $A$, $\mu_{tmp}$)
10        **if** $\mu_{tmp}$ is $feasible$ **and**
           $Obj(\mu_{tmp}, \mu_0) \leq Obj(\mu_b, \mu_0)$
11            $\mu_b := \mu_{tmp}$
12     **return** $\mu_b$

The BalCon sorts hosts by migration cost (see Definition 4) and attempts to release the hosts one by one (rows 3-4). To release host $h$, the procedure searches for a better feasible mapping $\mu_{tmp}$ in rows 5-9. In the better mapping, $h$ has to be turned off, and VMs from $h$ are reassigned to active hosts $A$. The VMs from $h$ are placed into the stash $S$ and unassigned from the mapping $\mu_{tmp}$ in rows 6-7. In rows 9-11 ForceFit procedure tries to fit VMs from the stash $S$ into active hosts $A$. If a new mapping $\mu_{tmp}$ is feasible and leads to an improvement of the objective function, then host $h$ is successfully released (rows 10-11).

The ForceFit heuristic is presented in Listing 2. In rows 2-3, we define two variables for the procedure: *ForceSteps* and *p*. The integer variable *ForceSteps* is used to control the number of iterations in the while loop (rows 4-19). To avoid coming back to previously-visited solutions, we prohibit a host from being selected more than $\gamma$ times in a row during Force Steps. The state information is maintained in the object instance *p*.

The ForceFit heuristic works until the stash is empty or the Force Steps limit is reached. In each step, ForceFit tries to insert the largest VM $v$ (Eq. 16) from the stash into the destination host $h$. The choice of the destination host depends on the cluster state, which is determined by the `Classify` procedure (rows 22-29). The procedure implements ideas from Section 4.2 and classifies the state into three classes: Ample (Section 4.4), Balanced (Section 4.5), and Lopsided (Section 4.6). The last two classes use Force Steps and differ in their approach to choose the destination host $h$ and VMs $V_e$ to eject from $h$.

The choice of VMs $V_e$ influences the number of induced migrations and the amount of migrated memory at the end of ForceFit. To reduce the amount of migrated memory, we prefer to eject VMs $V_e$ that were migrated to the destination host $h$ during previous steps of the algorithm. The shuffling of these

**Listing 2:** ForceFit heuristic

```
1  Procedure ForceFit(S, A, μ)
2      ForceSteps := 0
3      p := RepeatsProhibitor(A, γ)
4      while S is not empty and ForceSteps < b do
5          v := peek largest v in S
6          class := Classify(S, A, μ, v)
7          S := remove v from S
8          if class is "Ample"
9              μ := BestFit(v, A, μ)
10         if class is "Balanced"
11             ForceSteps +=1
12             h, p := ChooseHostBalanced(v, A, μ, p)
13             μ, V_e := ForceFitBalanced(v, h, μ)
14             S := add all VMs from V_e to stash S
15         if class is "Lopsided"
16             ForceSteps +=1
17             h, p := ChooseHostLopsided(v, A, μ, p)
18             μ, V_e := ForceFitLopsided(v, h, μ)
19             S := add all VMs from V_e to stash S
20     return μ
21
22 Procedure Classify(S, A, μ, v)
23     if ∃h ∈ A : v fits h
24         return "Ample"
25     cap := Capacity(S, A, μ)
26     pcap := PotentialCapacity(S, A, μ)
27     if cap < 1 or cap < α · pcap
28         return "Lopsided"
29     return "Balanced"
```

migrated VMs in ForceFit is unable to increase the objective function. In other words, only new migrations in ForceFit increase the objective function compared to the best step of BalCon $\mu_b$. In terms of memory, improving the objective function (row 10 of Listing 1) requires the amount of new memory for migration during ForceFit to be no more than $MPH$ ($M(\mu_{tmp}, \mu_0) - M(\mu_b, \mu_0) \leq MPH$).

*4.4. Ample class*

In the Ample class one or more hosts have enough free space to accommodate a VM. We employ a Best Fit heuristic to allocate VM $v$. The destination host $h$ is chosen based on surrogate load

$$|h| = \frac{\text{load}(h).cpu}{h.cpu} + \frac{\text{load}(h).mem}{h.mem}. \qquad (24)$$

Among all hosts with enough free space, we assign a VM to the host with the highest load.

*4.5. Balanced class*

To place a VM $v$ in the Balanced class (Listing 2 rows 11-14), some other VMs must be ejected and placed into the stash. The balanced situation is special because the majority of hosts are low in the same resource. Such scenarios makes the situation closer to a one-dimensional problem, where a smaller VM is easier to place than a larger one. Therefore, ChooseHostBalanced (row 12) chooses the destination host $h$ with the greatest number of VMs smaller than $v$ using Eq. 16.

Afterwards, ForceFitBalanced (row 13) sorts the VMs in $h$ according to lexicographical order:

1. Prefer VMs previously migrated to $h$.

2. In case of a tie, prefer VMs with smaller migration cost.

Next, ForceFitBalanced iterates over the sorted list of VMs and excludes them from $h$ one by one until there is enough free space in $h$ for $v$. After placing $v$ into $h$, the procedure tries to return the excluded VMs which fit into $h$ in reverse order. The remaining VMs are returned as $V_e$ (rows 15).

*4.6. Lopsided class*

In the Lopsided class (Listing 2 rows 16-19) we also need to eject some VMs from the destination host $h$ into the stash to place a given VM $v$. However, unlike the balanced case, both resources limit VMs placement. The Lopsided class is a key tool of the BalCon from a balancing point of view. The balance of the whole datacenter improves by the choices of the destination host and VMs to eject for future induced migrations.

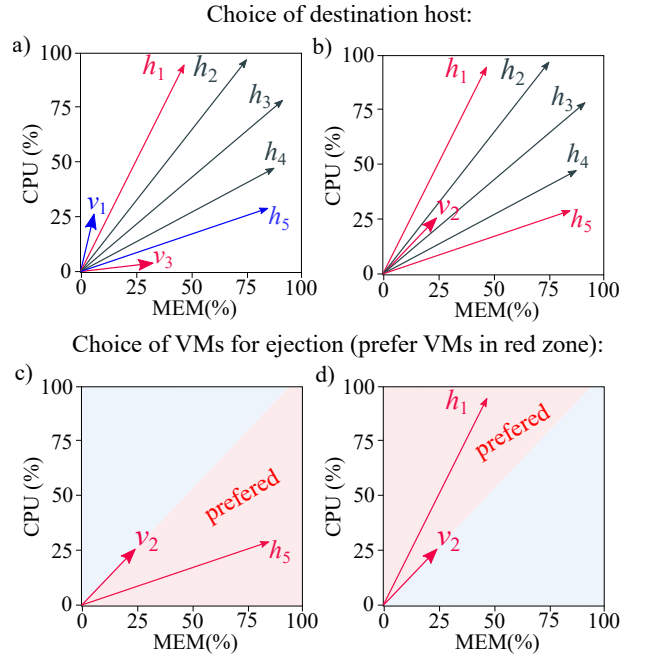

Figure 5: Illustration of choices of destination host and VMs for ejection in the Lopsided class. a-b) Illustrate the choice of the destination host ($h_1$, $h_5$) depending on the VM ($v_1$, $v_2$, $v_3$) to place. In a) the choice is explicit, and the destination host has the opposite load angle (for $v_1$ is $h_5$ and for $v_3$ is $h_1$), whereas in b) the choice is ambiguous and depends on the previous steps ($v_2$ can be allocated on $h_1$ and $h_5$). c-d) Illustrates choices of VMs for ejection. The preferable VMs are in the red areas that are located on the same side from the VM as the destination host ($h_1$ or $h_5$).

Figure 5 presents an example of the resource requirements of the VMs and the loads of the hosts. Some hosts lack CPU but have plenty of memory ($h_1$), some hosts lack memory but have plenty of CPU ($h_4$, $h_5$), and some are low on both resources. To reduce imbalance, we prefer to place a given VM into the host with minimal or maximal load angles (in example $h_1$ or $h_5$).

**Definition 7.** The load angle of a VM is $\arctan \frac{v.cpu}{v.mem}$, and the load angle of a host is $\arctan \frac{load(h).cpu}{load(h).mem}$.

To choose the destination host $h$, the procedure `ChooseHostLopsided` (row 17) uses the global variable $r \in \{"mem", "cpu"\}$ and follows rules:

1. If the load angle of $v$ is the largest or smallest among the load angles of all the hosts, choose the destination host with the most opposite load angle and set $r$ to the host's largest resource.
   For instance, in Figure 5a, $v_1$ has the largest angle and is moved to host $h_5$, which has the smallest angle ($r$ set to "mem"). Similarly, $v_3$ has the smallest angle and is moved to $h_1$, which has the largest angle ($r$ set to "cpu").

2. Else, switch $r$ and choose the host with the largest load by the resource in $r$.
   For instance, with $v_2$ from Figure 5b, if $r$ was "cpu", it switches to "mem", and $v_2$ is moved to $h_5$ (host with the largest "mem" load). Otherwise, it switches to "cpu" and is moved to $h_1$ (host with the largest "cpu" load).

The implementation of the `ForceFitLopsided` procedure (row 18) is similar to that of the Balanced class, except it has an additional requirements on the load angle of the VMs for ejection. To fix the destination hosts lopsidedness, the load angle of the VMs for ejection has to be close to the load angle of the host. The procedure follows lexicographical sorting of the VMs in $h$:

1. Prefer VMs with a direction to the same side of $v$ as $h$.
   For instance, if $h$ has a lower angle than $v$, we prefer VMs with angles less than $v$ (red zone Figure 5c). Otherwise, we prefer VMs with angle larger than $v$ (red zone Figure 5d).

2. In case of a tie, prefer VMs which were previously migrated to $h$.

3. In case of a second tie, prefer VMs with a smaller migration cost.

After sorting, the `ForceFitLopsided` procedure frees space in $h$ for $v$ exactly like in `ForceFitBalanced` and returns the VMs for ejection $V_e$ in row 18.

*4.7. Modified Sercon heuristics and time complexity*

To compare BalCon with well-known approaches, we modified the Sercon heuristic to fit our objectives. The SerconModified heuristic derives from BalCon if we forbid using Force Steps. The differences between the original Sercon heuristic and the SerconModified heuristic include:

1. In Sercon, the authors limited the total number of allowed migrations over all hosts. The SerconModified heuristic limits the amount of migrated memory in each algorithm step. A step is accepted if the new amount of migrated memory in the step is no more than *MPH* (Listing 1 row 10).

2. Unlike Sercon, we try to release each host once.

3. The destination host for a VM in the original version is chosen with First Fit, whereas we use Best Fit. Also, we choose the destination host among all active hosts.

4. We omitted the migration efficiency parameter.

The second modification leads to $|H|$ times less complexity of SerconModified compared to SerconOriginal (see Table 2), where $V_{max}$ is the maximum number of VMs in a host. The complexity of BalCon is only $b$ times worse than that of SerconModified. The upper bounds on $V_{max}$ can be $|V|$.

Table 2: Comparison of time complexity of the algorithms.

| Algorithm | Time complexity |
| --- | --- |
| SerconOriginal | $O(|H|^2 \cdot (V_{max} \cdot \log(V_{max}) + V_{max} \cdot |H|))$ |
| SerconModified | $O(|H| \cdot (V_{max} \cdot \log(V_{max}) + V_{max} \cdot |H|))$ |
| BalCon | $O(b \cdot |H| \cdot (V_{max} \cdot \log(V_{max}) + V_{max} \cdot |H|))$ |

## 5. Evaluation

*5.1. Datasets*

Algorithm evaluation was performed using Huawei and synthetic datasets. Synthetic datasets was generated to test the algorithms under complex inputs, whereas the Huawei datasets validate algorithm performance on real data. The comparison of the main characteristics of the datasets is presented in Table 3 and Firgure 6. In general, synthetic problem instances are less balanced, have more VMs and more flavors than the Huawei datasets.

Table 3: Statistics of Huawei and synthetic datasets, where mean values are given per instance.

| Dataset | Instances Total | Hosts Mean | VMs Mean | Flavors Mean/Total | Balance factor Mean |
| --- | --- | --- | --- | --- | --- |
| Huawei | 555 | 96 | 1.0k | 8 / 69 | 0.99 |
| Synthetic | 200 | 50 | 2.7k | 28 / 30 | 0.07 |

The Huawei dataset contains 555 cluster snapshots from an operational Huawei Cloud. The snapshots were gathered from clusters of various sizes and roles. The flavor distribution over all snapshots is given in Figure 6a, whereas the distribution of hosts within a range of the number of VMs is presented in Figure 6c.

To generate a complex synthetic dateset, we created instances with severe lopsidedness of resources. Briefly, the VMs were generated from 30 flavors with decreasing probabilities by CPU size (see Figure 6b). Then, the VMs were sorted by the load angle (Def. 7) and packed into hosts using the First Fit heuristic. Such packing method complicates the ability of Sercon-like heuristic because of the lack of free space in hosts. Also, the packing produces an imbalance that requires many Force Steps in BalCon to place VMs with ForceFit. To validate the lopsidedness of resources in our synthetic datasets we calculated the mean balance factor. The balance factor of an instance from the datasets was calculated with Eq. 19, where $s = \frac{1}{|H|} \sum_{h \in H}(h.cpu, h.mem)$ is the mean hosts capacity in the instance. The mean balance factor (Table 3) of the initial feasible mappings in the Huawei instances is 0.99, which corresponds to a balanced situation. In contrast, for the synthetic data, the mean balance factor is 0.07 (Table 3), indicating a highly lopsided situation.

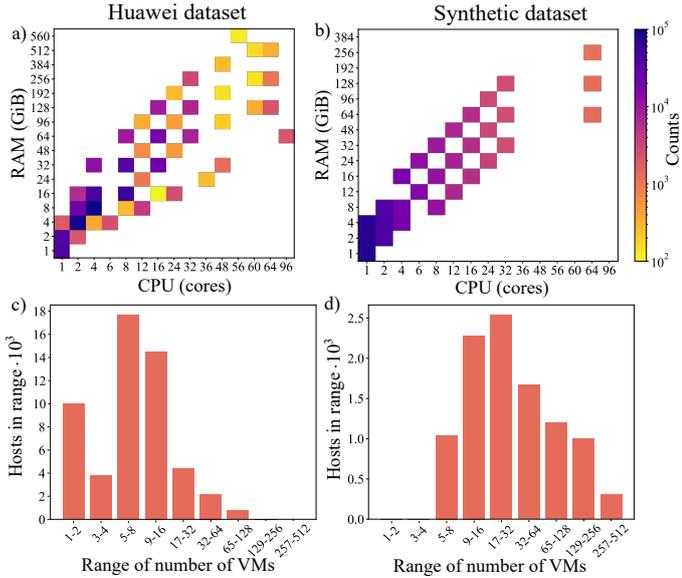

Figure 6: The comparison between the Huawei (a,c) and synthetic (b,d) datasets. Panels a-b present the flavor distribution over all problem instances, whereas panels c-d) demonstrate the number of hosts with VMs in the given range.

*5.2. BalCon performance*

We begin evaluation of algorithm performance with the extreme case of the classical VBP problem, where the maximal memory for migration per host (Eq. 2) is infinite $MPH = \infty$ TiB. To compare BalCon with SerconModified (Section 4.7), we measured the gap between the values of the objective function of the algorithm solution $\mu_{alg}$ and the optimal solution $\mu_{opt}$

$$Gap(\mu_{alg}, \mu_{opt}) = \frac{Obj(\mu_{alg}, \mu_0) - Obj(\mu_{opt}, \mu_0)}{Obj(\mu_0, \mu_0) - Obj(\mu_{opt}, \mu_0)}, \quad (25)$$

where $\mu_{opt}$ were obtained with the Flavor flow model (Section 3.2). The performance profile [37] indicates the advantage of BalCon over SerconModified, which optimally solved 535 and 345 instances respectively on Huawei dataset (see Figure 7a). Also, the original version of Sercon (see Section 4.7) optimally solved 245 instances. On the synthetic datasets BalCon optimally solved all the problem instances, whereas Sercon was unable to solve any because of lack of free space of one of the resources (see Figure 7b and Section 5.1). The iterative nature of the algorithms allows for easy implementation of a running time limit. The time limit for all algorithms was set to 60 seconds on one physical core of Intel Xeon E5-2690. However, on average, BalCon solved instances in less time (see Table 4). Additionally, the measured execution time is well-correlated with the algorithms time complexity from Table 2.

Table 4: The comparison of execution time between methods in the case of the VBP problem and Huawei dataset.

| Method | ILP Flow Model | Relaxed ILP Flow model (LB) | BalCon | Sercon Original | Sercon Modified |
|---|---|---|---|---|---|
| Mean execution time | 4h | 0.3s | 9.5s | 0.3s | 0.1s |

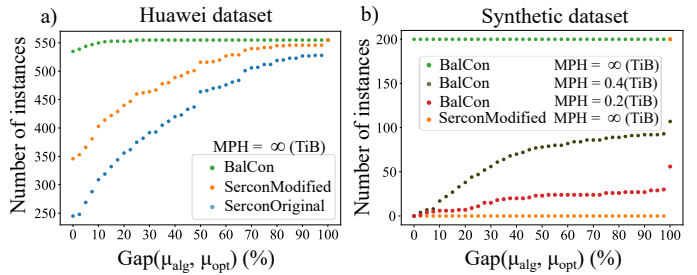

Figure 7: The performance profiles of BalCon, SerconModified, and SerconOriginal are compared to the optimal solution $\mu_{opt}$ of the Flavor flow model for the VBP problem when $MPH = \infty$ (TiB). The $Gap$ is calculated using Eq. 25. Panels a) and b) show the comparison for the Huawei and Synthetic datasets, respectively.

To investigate the algorithms performance towards migration-aware consolidation, we chose a few values of $MPH$. Solvers with the Flavor flow model were unable to obtain optimal solutions for all $MPH$ values in a reasonable time. Therefore, we use the LB obtained with the relaxed flavor flow model instead of the true optimum (Section 3.3). In the example of the VBP problem the execution time for the LB is significantly lower (see Table 4). The mean values of $Gap(\mu_{alg}, \mu_{LB})$ over non-trivial solutions – at least one host is released – are presented in Figure 8. As expected from the VBP results, BalCon outperforms SerconModified on the Huawei dataset and dominates on the Synthetic dataset, where SerconModified is unable to solve any instance. The Force Steps of BalCon allow to provide closer solutions to the LB than SerconModified on the Huawei dataset (Figure 8a) at large values of $MPH$ from 1 TiB to 10 TiB. However, when $MPH$ reaches the capacities of the hosts $\sim 0.7$TiB, the number of non-trivial instances decreases, and performance of both algorithms equalizes, due to the insufficient memory for induced migrations. Even at $MPH = 0.2$ TiB, Force Steps lead to a little bit worse performance of BalCon. Qualitatively BalCon demonstrates similar operation in the Synthetic datasets, however the $Gap$ to LB is larger and the number of non-trivial solutions is more sensitive to $MPH$. The last facts are because of the high imbalance of the Synthetic datasets, which require additional amounts of memory for migration.

In general, BalCon's high performance is determined by the Ample, Balanced, and Lopsided classes and their respective heuristics (see Sections 4.4 - 4.6). Other factors – such as cost functions for VMs ordering – are details of minor importance. For instance, a random ordering of VMs in the stash leads to less than 1% of Gap increase compared to ordering with Eq. 16. To demonstrate the influence of *MPH* on BalCon's performance and areas for further improvement of the algorithm, we selected one representative instance from each dataset (see Figure 9). The number of active hosts and total migrated memory of SerconModified and BalCon coincide at low *MPH* (Figure 9a). Then, at larger *MPH*, SerconModified reaches constant values, whereas BalCon releases more hosts at the cost of a larger amount of migrated memory and improved objective function.

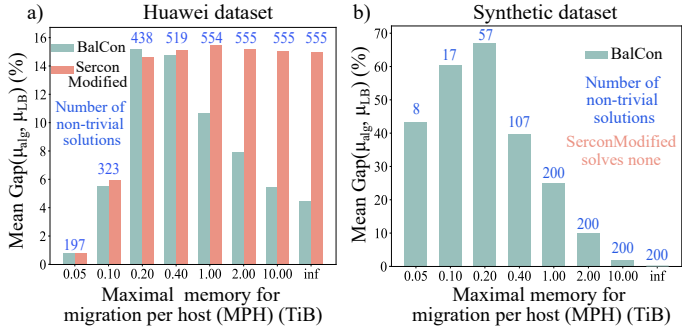

Figure 8: Comparison of BalCon and SerconModified with lower bound (LB) at different values of *MPH*. The mean *Gap* is calculated over non-trivial solutions using Eq. 25. Panels a) and b) show the results on the Huawei dataset and Synthetic dataset, respectively.

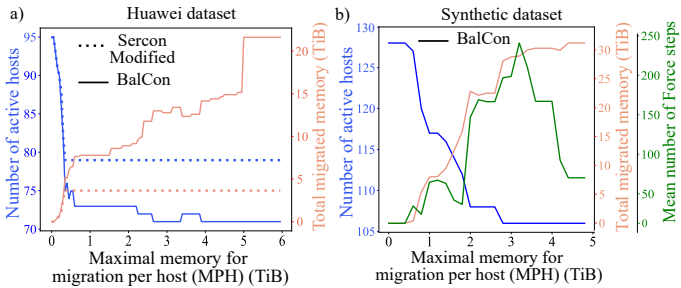

Figure 9: Examples of BalCon and SerconModified performance at different *MPH* values. The evaluation is demonstrated in terms of the number of active hosts, the amount of migrated memory, and the number of Force Steps. The problem instances are from a) Huawei dataset and b) Synthetic dataset.

Because of the greedy nature, BalCon sometimes performs non-optimal steps by increasing the amount of migrated memory, the number of active hosts, or the number of Force Steps. For instance, the number of active hosts is increased by one in the range of [3.4,3.8] TiB, which could be avoided using the solution at 3 TiB (see Figure 9a). Also, at the same number of active hosts, the algorithm uses larger memory than it could in the range [5.0, 6.0] TiB because the better solution is at 4 TiB. The mean number of Force Steps changes, although the number of active hosts remains the same for the range [2.8, 3.6] TiB (see Figure 9b). We would like to emphasize that those are only examples of non-optimal BalCon operations, which might be considered for further algorithm improvement. In general, the number of Force Steps and total migrated memory increase along with *MPH*. Also, on average BalCon demonstrates outstanding performance, especially in the case of large *MPH* and imbalanced situations (Figure 8).

## 6. Conclusions

We proposed the BalCon algorithm which efficiently solves the migration-aware consolidation problems. The algorithm was compared with a modified Sercon heuristic and ILP models. The advantages of BalCon over Sercon-like heuristics were achieved due to Force Steps. The Force Steps allow BalCon to optimally solve imbalanced problem instances that Sercon-like heuristics are unable to optimize. The performance of BalCon is very close to optimal at large values of *MPH*. Time complexity of BalCon is only $b$ times larger than that of modified Sercon heuristic, where $b$ is the maximum number of Force Steps. Note that the BalCon implementation is independent of flavor set $F$ and therefore directly applicable to dynamic consolidation. Also, we used the amount of RAM as a migration cost, however one can replace it with other metrics such as number of VM migrations, predicted time for migration of VM, probability of SLA violation, *etc*.

## 7. Data availability

The data and code used in this article are available by the link https://github.com/andreigudkov/BalCon.

## 8. Declaration of competing interest

The authors declare that they have no known competing financial interests or personal relationships that could have appeared to influence the work reported in this paper.

## 9. Acknowledgments

Authors are gratefull to Mr. Oliver Slumbers for helping with language polishing of this article. Dr. Stepan Romanov thanks Dr. Mafuda for fruitfull discussions and organizational help.

## 10. CRediT authorship contribution statement

The authors equally contributed to this work.